\documentclass[
    ,final            
  ]
  {aipproc}

\layoutstyle{6x9}

\usepackage{eurosym}

\begin{document}

\title{Altruistic behavior pays, or the importance of fluctuations
in evolutionary game theory}

\author{Angel S\'anchez}{
  address={Grupo Interdisciplinar de Sistemas Complejos (GISC),
Departamento de Matem\'aticas, Universidad Carlos III de Madrid,
28911 Legan\'es, Madrid, Spain}
}

\author{Jos\'e A.\ Cuesta}{
  address={Grupo Interdisciplinar de Sistemas Complejos (GISC),
Departamento de Matem\'aticas, Universidad Carlos III de Madrid,
28911 Legan\'es, Madrid, Spain}
}

\author{Carlos P.\ Roca}{
  address={Grupo Interdisciplinar de Sistemas Complejos (GISC),
Departamento de Matem\'aticas, Universidad Carlos III de Madrid,
28911 Legan\'es, Madrid, Spain}
  ,altaddress={Centro Nacional de Biotecnolog\'\i a, 28049 
Cantoblanco, Madrid, Spain} 
}

\begin{abstract}
Human behavior is one of the main problems for evolution, as it is 
often the case that human actions are disadvantageous for the self
and advantageous for other people. Behind this puzzle are our beliefs
about rational behavior, based on game theory. Here we show that by
going beyond the standard game-theoretical conventions, apparently 
altruistic behavior can be understood as self-interested. We discuss 
in detail an example related to the so called Ultimatum game and 
illustrate the appearance of altruistic behavior induced by fluctuations.
In addition, we claim that in general settings, fluctuations play a very
relevant role, and we support this claim by considering a completely 
different example, namely the Stag-Hunt game.
\end{abstract}

\maketitle


\section{Introduction}

Human altruistic behavior is a long-standing problem in 
evolutionary theory, as first realized by Darwin himself:
\begin{quotation}
He who was ready to sacrifice his life (\dots) rather than betray his comrades,
would often leave no offspring to inherit his noble nature\dots Therefore,
it seems scarcely possible (\dots) that the number of men gifted with such virtues
(\dots) would be increased by natural selection, that is, by the survival of
the fittest. 
\cite{Darwin-Descent}
\end{quotation}
At the crux of the problem lies the fact that Darwin developed his 
theory assuming that natural selection acts exclusively on individuals 
only. On this grounds, he could not possibly understand altruistic behavior
in humans, i.e., acts that decrease the actor's fitness while increasing that
of others. Reluctantly, he had to call for selection at group level:
\begin{quotation}
A man who was not impelled by any deep, instinctive feeling, to sacrifice
his life for the good of others, yet was roused to such actions by a sense
of glory, would by his example excite the same wish for glory in other men,
and would strengthen by exercise the noble feeling of admiration. He might
thus do far more good to his tribe than by begetting offsprings with a
tendency to inherit his own high character.
\cite{Darwin-Descent}
\end{quotation}
In fact, human behavior is unique in nature. Indeed, altruism or 
cooperative behavior exists in other species, but it can be 
understood in terms of genetic relatedness (kin selection,
introduced by Hamilton \cite{Hamilton:1964}) or of repeated 
interactions (as proposed by Trivers \cite{Trivers:1971}).
However, human cooperation extends to genetically unrelated 
individuals and to large groups, characteristics that cannot 
be understood within those schemes. Subsequently, a number of 
theories based on group and/or cultural evolution have been put 
forward in order to explain altruism (see
\cite{Hammerstein} for a review). 

\section{The Ultimatum game}

In order to address quantitatively the issues above, behavioral
researchers use evolutionary game theory \cite{Gintisbook,Camerer} to 
design experiments that try to find the influence of different
factors. 
In this paper, we analyze this problem in the context of a 
specific set of such experiments, related to the so called 
Ultimatum game \cite{Guth:1982,Henrich}. 
In the Ultimatum game,
under conditions of anonymity, two players are shown a sum
of money, say 100 \EUR{}. One of the players, the ``proposer'',
is instructed to offer any amount, from 1 \EUR{} to 100 \EUR{},
to the other, the ``responder''. The proposer can make only one
offer, which the responder can accept or reject. If the offer is
accepted, the money is shared accordingly; if rejected, both
players receive nothing. Since the game is played only once
(no repeated interactions) and anonymously (no reputation gain;
for more on explanations of altruism relying on reputation 
see \cite{Nowak-Sigmund}),
a self-interested responder will accept any amount of money
offered. Therefore, self-interested proposers will offer the
minimum possible amount, 1 \EUR{}, which will be accepted.
Notwithstanding,
in actual Ultimatum game experiments with human subjects,
average offers do not even approximate the self-interested prediction.
Generally speaking, proposers offer respondents very substantial
amounts (50 \% being a typical modal offer) and respondents
frequently reject offers below 30 \% \cite{Fehr2003}. Most of the
experiments have been carried out with university students in
western countries, showing a large degree of individual variability
but a striking uniformity between groups in average behavior.
A large study in 15 small-scale societies \cite{Henrich}
found that, in all
cases, respondents or proposers behave in a reciprocal manner.
Furthermore, the behavioral variability across groups was much
larger than previously observed: while mean offers in the case
of university students are in the range 43\%-48\%, in the
cross-cultural study they ranged from 26\% to 58\%.

The fact that indirect reciprocity is excluded by the anonymity
condition and that interactions
are one-shot (i.e., repeated interaction does not apply) 
allows one to interpret rejections in terms of the so-called
strong reciprocity \cite{Gintis2000,Fehr2002}.
This amounts to considering that
these behaviors are truly altruistic, i.e., that
they are costly for the individual performing them in so far as
they do not result in direct or indirect benefit. As a consequence,
we return to our evolutionary puzzle: The negative effects of
altruistic acts must decrease the altruist's fitness as compared to
that of the recipients of the benefit, ultimately leading to
the extinction of altruists. Indeed, standard evolutionary game
theory arguments applied to the Ultimatum game lead to the expectation
that in a mixed population, punishers (individuals
who reject low offers) have less chance to survive
than rational
players (indivuals who accept any offer) and eventually disappear.
In the remainder of the paper, we will show that this conclusion 
depends on the dynamics, and that different dynamics leads to 
the survival of punishers through fluctuations. 

\section{The model}

We consider a population of $N$ players (agents) of the Ultimatum game
with a fixed sum of money $M$ per game.
Random pairs of players are chosen, of which one is the proposer
and another one is the respondent. In its simplest version,
we will assume that
players are capable of other-regarding behavior (empathy); consequently,
in order to optimize their gain,
proposers offer the minimum amount of money
that they would accept. Every agent has her own, fixed
acceptance threshold, $1\leq t_i\leq M$ ($t_i$ are always integer
numbers for simplicity). Agents have only one strategy:
respondents reject any offer
smaller than their own acceptance threshold, and
accept offers otherwise.
Money
shared as a consequence of accepted offers accumulates to the
capital of each of the involved players. As our main aim is to
study selection acting on modified descendants, hereafter we interpret this
capital as `fitness'
(here used in a loose, Darwinian sense, not in the more
restrictive one of reproductive rate).
After $s$ games,
the agent with the overall minimum fitness is
removed (randomly picked if there are several)
and a new agent is introduced by duplicating that
with the maximum fitness, i.e., with the same threshold and the
same fitness (again randomly picked if there are
several). Mutation is introduced in the duplication process by
allowing changes of $\pm 1$ in the acceptance threshold of the
newly generated player with probability 1/3 each. Agents
have no memory (i.e., interactions are one-shot) and no information
about other agents (i.e., no reputation gains are possible).
We stress that the model is dramatically simplified; however, we 
have studied more complicated versions (including separate acceptance
and offer thresholds) and the results are similar to the ones we 
discuss below. Another factor we have considered is smaller mutation
rates, again without qualitative changes in the result. Therefore, 
for the sake of brevity we concentrate here on the simple model 
summarized above and refer the reader to \cite{Cuesta-Sanchez}
for a more detailed analysis including those other versions.

\section{Results}

\begin{figure}
  \label{figure1}
  \includegraphics[height=.2\textheight]{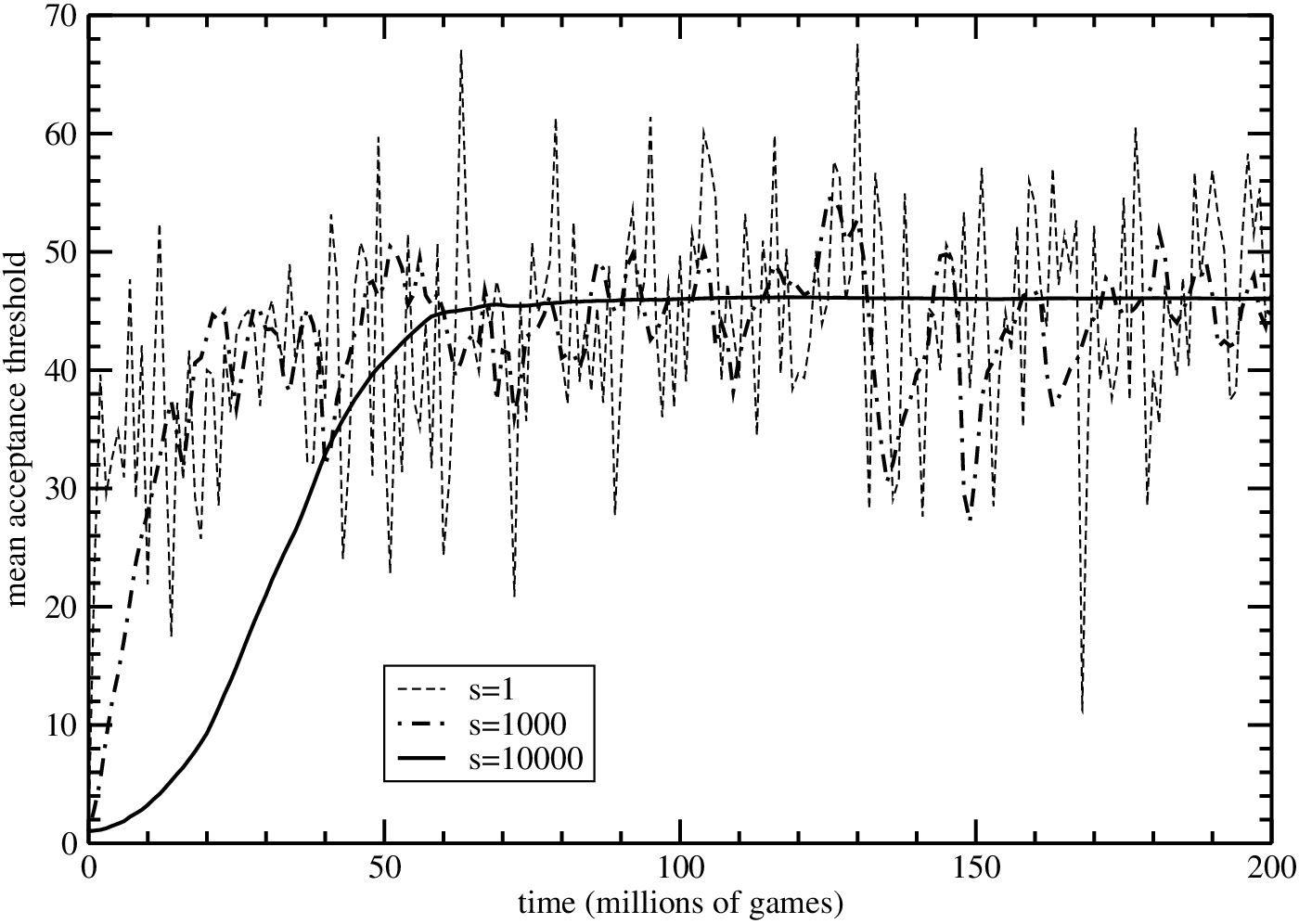}
  \hspace*{5mm}
  \includegraphics[height=.2\textheight]{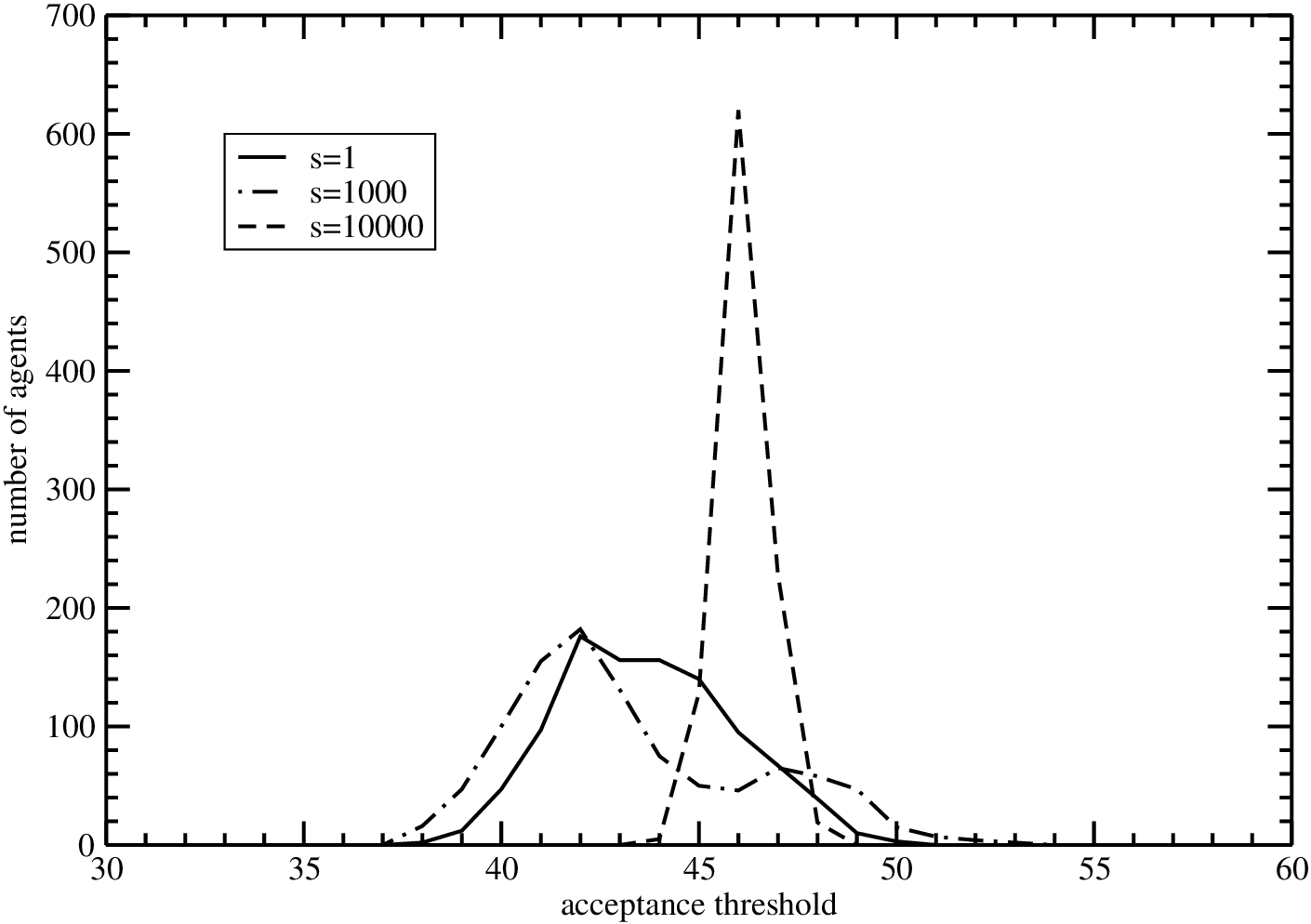}
  \caption{Left: mean acceptance threshold as a function of simulation 
  time. Initial condition is that all agents have $t_i=1$.
  Right: acceptance threshold distribution after $10^8$ games. 
  Initial condition is that all agents have uniformly distributed, random $t_i$.
  In both cases, $s$ is as indicated from the plot.}
\end{figure}

Figure \ref{figure1} shows the typical outcome of simulations of our model.
As we can see, the mean acceptance threshold rapidly evolves towards 
values around 40\%, while the whole 
distribution of thresholds converges to a peaked function, with
the range of acceptance thresholds for the agents covering about a
10\% of the available ones.
These are values compatible with the experimental results discussed 
above. The 
mean acceptance threshold fluctuates
during the length of the simulation, never reaching a stationary value
for the durations we have explored. The width of the peak fluctuates
as well, but in a much smaller scale than the position.
The fluctuations are larger for smaller values of $s$, and when $s$
becomes of the order of $N$ or larger, the evolution of the mean 
acceptance threshold is very smooth. This is a crucial point and will 
be discussed in more detail below.  
Importantly, the typical evolution
we are describing does not depend on the initial condition. In particular,
a population consisting solely of self-interested agents, i.e., all
initial thresholds are set to $t_i=1$, evolves in the same fashion.
Indeed, the distributions shown in the left panel of Figure 
\ref{figure1} have been obtained with such an initial condition, 
and it can be clearly observed that self-interested agents disappear
in the early stages of the evolution. 
The number of players and the value $M$ of the capital at stake in every
game are not important either, and increasing $M$ only leads
to a higher resolution of the threshold distribution function.

\section{Discussion}

As we mentioned in the preceding section, 
we have observed that taking very large values for $s$ or, strictly
speaking, considering the limit $s/N\to\infty$, does lead to different
results. In this respect, let us recall
previous studies of the Ultimatum game by Page and
Nowak \cite{PageNowak00,PageNowak02}.
The model introduced in those works has a dynamics completely
different from ours: following standard evolutionary game theory,
every player plays every other one in both roles (proponent and
respondent), and afterwards players reproduce with probability
proportional to their payoff (which is fitness in the reproductive
sense). Simulations and adaptive dynamics equations show then that the
population ends up composed by players with fair (50\%) thresholds.
This is different from our observations, in which we hardly ever
reach an equilibrium (only for large $s$) and even then equilibria
set up at values different from the fair share. The reason for this
difference is that the Page-Nowak model dynamics describes the
$s/N\to\infty$ limit of our model, in which between death-reproduction
events the time average gain all players obtain is
the mean payoff with high accuracy.
We thus see that our model is more general
because it has one free parameter, $s$, that allows selecting different
regimes whereas the Page-Nowak dynamics is only one limiting case.
Those different regimes are what we have described as fluctuation dominated
(when $s/N$ is finite and not too large) and the regime analyzed by
Page and Nowak (when $s/N\to\infty$).
This amounts to saying that by varying $s$ we can
study regimes far from the standard evolutionary game theory
limit. As a result, we find a variability of outcomes for the
acceptance threshold consistent with the observations in real
human societies %
\cite{Henrich,Fehr2003}.

In fact, fluctuations due to the finite number of games are at the 
heart of our results. Among the results summarized above, the
evolution of a population entirely
formed by self-interested players into a diversified population with a
large majority of altruists is the most relevant and surprising one.
We will now argue that the underlying reason for this is precisely 
the presence of
fluctuations in our model. For the sake of definiteness, let us
consider the case $s=1$ (agent replacement takes place after every game)
although the discussion applies to larger (but finite) values of $s$ as
well. After one or more games, a mutation event will take place
and a ``weak altruistic punisher'' (an agent with $t_i=2$) will appear
in the population,
with a fitness inherited from its ancestor. For this new agent to be
removed at the next iteration so that the population reverts to its
uniform state, our model rules imply that this agent has to have
the lowest fitness, that is the only one with that value of fitness,
{\em and also} that it does not play as a proposer in
the next game (if playing as a responder the agent will earn nothing
because of her threshold). In any other event this altruistic punisher
will survive at
least one cycle, in which an additional one can appear by mutation.
It is thus clear that fluctuations indeed help altruists to take 
over: As soon as a few altruists are present in the population, it is
easy to see analytically that they will survive and proliferate even
in the limit $s/N\to\infty$.

\section{The Stag-Hunt game}

This far, we have shown that considering that players play a finite 
number of games between death-birth events in the Ultimatum game leads
to results unexpected from standard evolutionary game theory arguments. 
Hence, the question arises as to whether this is a consequence of the
many strategies available in the Ultimatum game (as many as possible 
values for $t_i$, 100 with our choice for the parameters) or, on the
contrary, it is a general phenomenon. To show that the latter is the
case, we have considered a completely different, much simpler kind of
game: the so-called Stag-Hunt game \cite{Gintisbook,Camerer,Henrich}. 
In this game, two hunters cooperate in hunting for stag, which is the
most profitable option;
however, hunting a stag is impossible unless both work together, and
they have the option of hunting for rabbit, less profitable, but 
with sure earnings. This is reflected in the following payoff matrix 
(C stands for cooperation in hunting stag, D stands for defection and 
hunting rabbit alone):
\begin{center}
\begin{tabular}{|c||c|c|}
\hline
\mbox{ } & C & D \\ \hline
\hline
C & 6 & 0 \\ \hline
D & 5 & 1 \\ \hline
\end{tabular}
\end{center}
This game belongs in the class of coordination games: In the 
language of game theory, it has two 
Nash equilibria, (C,C) and (D,D), and the players would like 
to coordinate in choosing the first one (so called payoff-dominant).
However, the second one is a safer choice 
because  it has the largest guaranteed minimum payoff 
(so called risk-dominant). 

We have been working on the evolutionary dynamics of this game and,
specifically, on the equilibrium selection problem 
\cite{todos}. For this example, we have chosen the dynamics given 
by the Moran process \cite{Nowak2004}, in which after $s$ games
an agent is duplicated with probability proportional to the fitness
accumulated during the $s$ games, and another one is killed 
randomly. With such a simple dynamics, it is an elementary 
exercise to show that, in the limit $s/N\to\infty$, the whole 
population becomes C (resp.\ D) strategists if the initial density 
of C strategists is larger (resp.\ smaller) than 1/2. As Fig.\ 
\ref{figure2} shows, simulation results for finite
$s$ are largely different from that analytical prediction:
\begin{figure}
  \label{figure2}
  \includegraphics[height=.2\textheight]{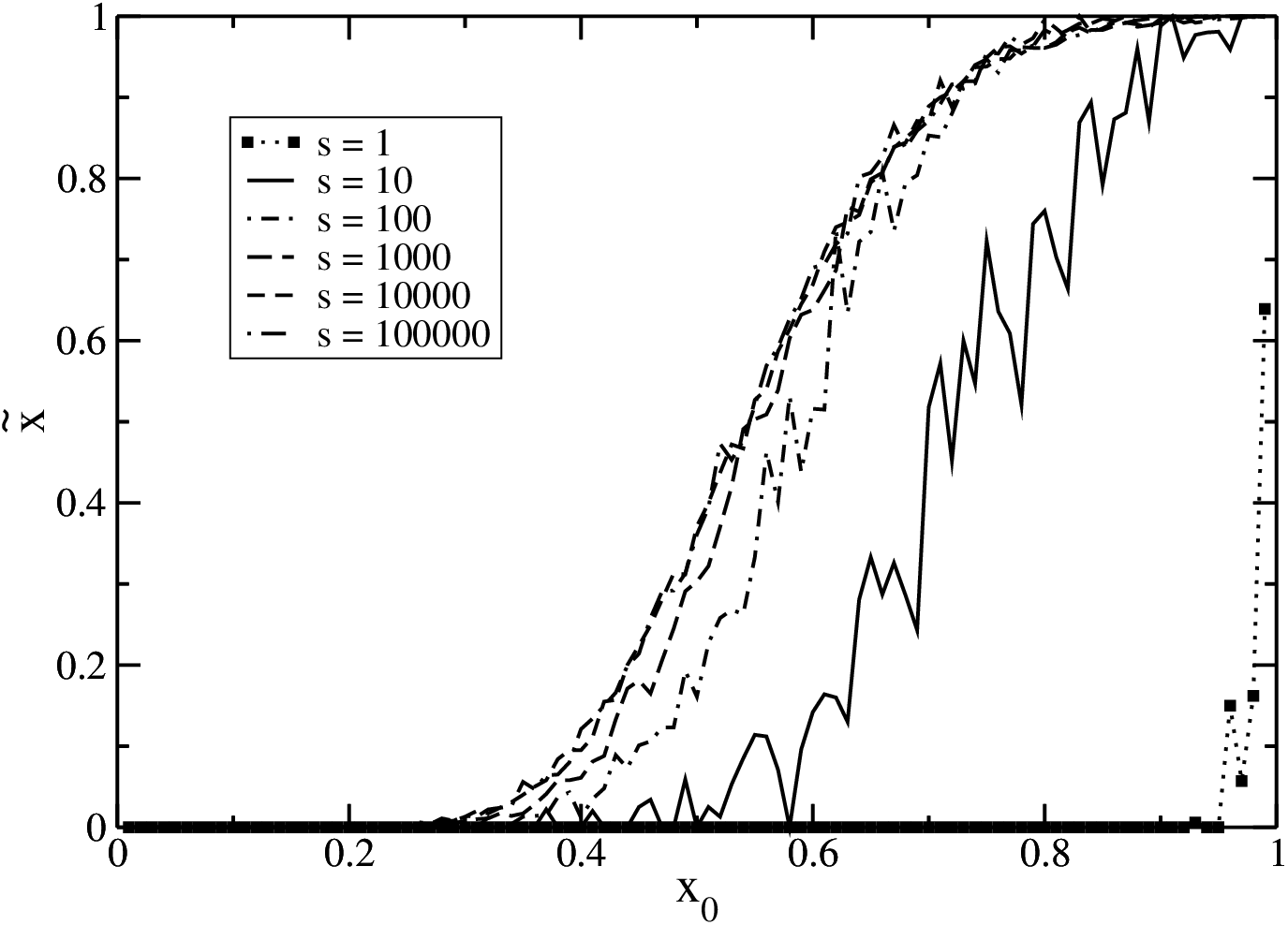}
  \hspace*{5mm}
  \includegraphics[height=.2\textheight]{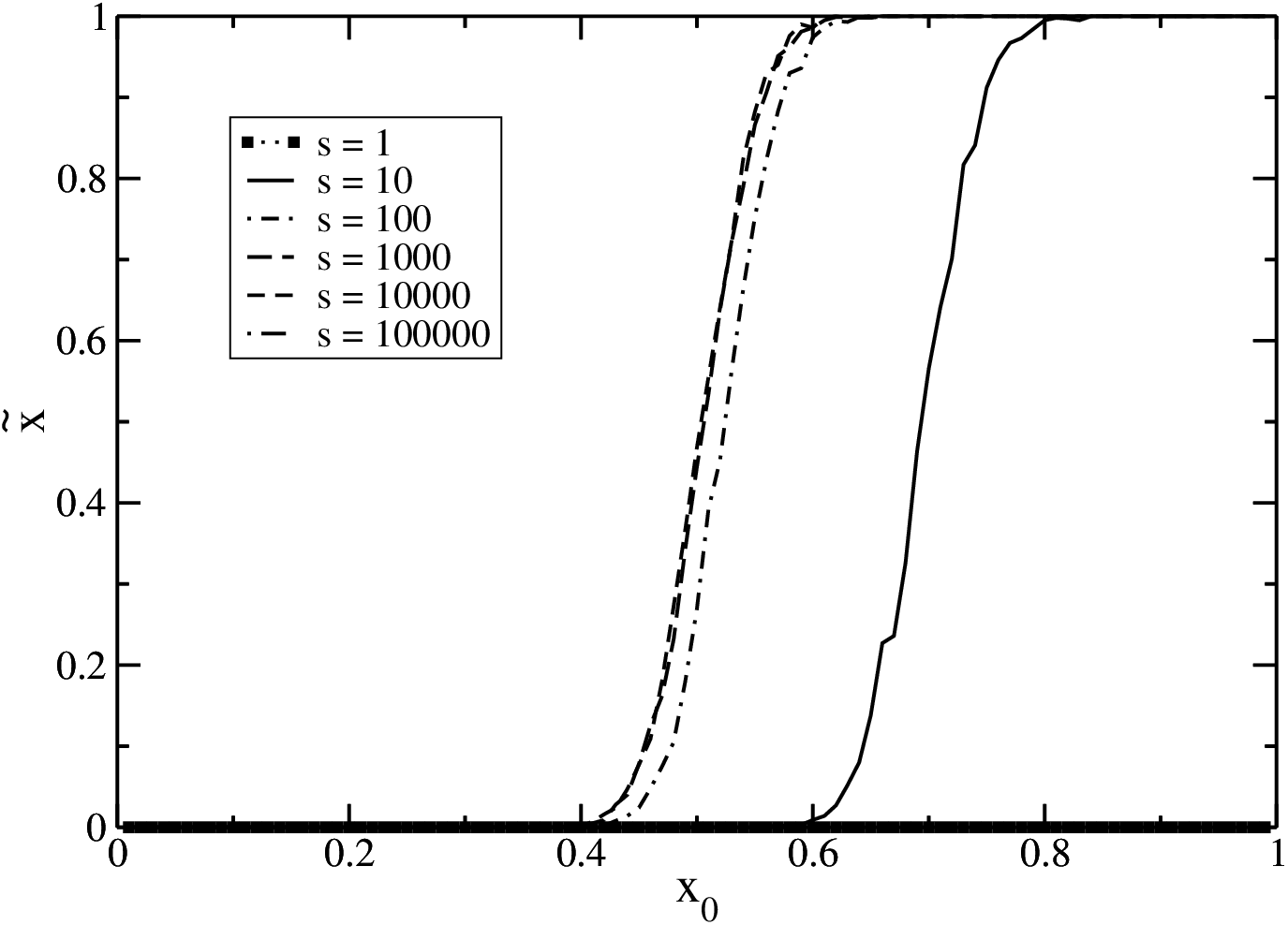}
  \caption{Fraction of games that end up with a cooperator-only
   population vs density of cooperators in the initial state
   for $N=100$ (left) and $N=1000$ (right) agents playing 
   the Stag-Hunt game. Results are obtained from simulations of the Stag-Hunt 
   game with the Moran dynamics, and for every initial density the final 
   density is averaged over 100 games. Values of $s$ are as indicated in the plot.}
\end{figure}
Indeed, we see that for cooperators to prevail in the final state, an initial
density larger than 1/2 is needed. In particular, for $s=1$, all agents become
defectors except for initial densities close to 1 in the case $N=100$ (left 
panel), and for
all initial densities for $N=1000$ (right panel) or larger (not shown). The 
plots also show that larger populations lead to better statistics (meaning that
curves are smoother and less noisy; it is evident that $\tilde{x}$ has a smaller
variance for larger populations), and the 
trend upon increasing $N$ is that the curves become step functions (as should
be for an infinite population). Importantly, the effect, namely that the basin of
attraction of the (D,D) equilibrium is enlarged for finite $s$, persists even
in the infinite population limit. In addition, it is also robust upon changes 
in the dynamics: we have verified that choosing the agent to be eliminated with
probability inversely proportional to the agent's fitness leads to qualitatively
similar result. We are thus faced with another clear-cut manifestation of the
relevance of taking the limit of infinite games before the dynamics occurs or,
on the contrary, sticking to a finite number of games. Once again, we stress that
the setup is completely different from the Ultimatum game and, as a consequence,
we claim that this kind of phenomena is generic and should be observed in 
many other problems. 

\section{Conclusions}

In this paper, we have shown that altruistic-like behavior, specifically,
altruistic punishment, may arise by means of exclusive individual selection
even in the absence of repeated interactions and reputation gains. Our
conclusion is important in so far as it is generally believed that
some kind of group selection is needed to understand the observed human
behavior. The reason for that is that game theoretical arguments
apparently show that altruists are at disadvantage with respect to
selfish individual. In this respect, another relevant conclusion
of the present work is that
perspectives and approaches alternative to standard evolutionary
game theory may be needed in order to understand paradoxical
features such as the appearance of altruistic punishment.
As additional evidence supporting this claim,
we have briefly discussed, in the context of the much simpler problem
of the stag-hunt game, that 
equilibrium selection is indeed dramatically modified by taking 
into account a finite number of games. Therefore, we conclude that 
the dynamics postulated for a particular application of evolutionary 
game theory must be closely related to the specific problem as the 
outcome can be completely different depending on the dynamics.

\begin{theacknowledgments}
AS thanks the organizers of the 8th Granada Seminar, specially Joaqu\'\i n
Marro, for the opportunity to present these results and to discuss with 
the Seminar attendees. 
We acknowledge financial support from Ministerio de Ciencia y Tecnolog\'\i a
(Spain) through grants BFM2003-07749-C05-01 (AS) and BFM2003-0180 (JAC).
\end{theacknowledgments}


\bibliographystyle{aipproc}   

\bibliography{granada05}

\IfFileExists{\jobname.bbl}{}
 {\typeout{}
  \typeout{******************************************}
  \typeout{** Please run "bibtex \jobname" to optain}
  \typeout{** the bibliography and then re-run LaTeX}
  \typeout{** twice to fix the references!}
  \typeout{******************************************}
  \typeout{}
 }

\end{document}